\documentclass[12pt,oneside]{article}
\usepackage{graphics}
\usepackage{epsfig}
\newcommand{\keg}[1]{$K^{{#1}} \rightarrow \pi e \nu \gamma$}
\begin{document}
  
  \def\saltaunrigo{\vspace{6pt}}
  \begin{center}
    \vspace{1.3cm}
    
	   {\Large \bf Monte Carlo simulation for radiative kaon decays}\\

\vspace{0.5cm}
  \vspace{1.3cm}
    {\bf C.~Gatti} \\
    \saltaunrigo
    {\em Laboratori Nazionali di Frascati - INFN} \\
    \saltaunrigo
    {\em claudio.gatti@lnf.infn.it} \\
  \vspace{0.9cm}
\normalsize  {\bf Abstract}
\end{center}
\small
For high precision measurements of $K$ decays, 
the presence of radiated photons cannot be neglected. 
The Monte Carlo simulations must include the radiative corrections
in order to compute the correct event counting and efficiency
calculations. In this paper we briefly describe a 
method for simulating such decays.
  
\normalsize
\clearpage
\section{Introduction}

Many measurements on $K$ decays have reached a statistical 
error close to 1\% or less. With such precision, the
presence of radiated photons, and in general the effect of 
radiative corrections, cannot be neglected.
Furthermore, the treatment of
the radiative corrections is explicitly required
for the extraction of many physical quantities, 
such as the CKM matrix element $V_{us}$ and 
the phase shifts $\delta_{0}-\delta_{2}$,
at precisions of a few percent (or above).
Hence, it is mandatory to include the effect of radiated photons 
in the Monte Carlo (MC) simulations.  

\begin{figure}[htbp]
  \begin{center}
    \includegraphics[totalheight=7.cm]{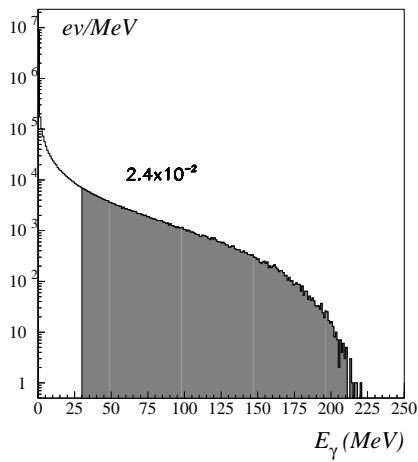}
    \caption{Energy spectrum for \keg{0}\ MC decays. 
      2.4\% of the events have $E_{\gamma}>30~\mbox{MeV}$.}
    \label{introspettro}
  \end{center}
\end{figure} 

There are two aspects of a measurement that are affected by
the presence of radiated photons~\cite{radcorr}:
the geometrical acceptance and the counting of the events.
About 2.4\% of \keg{0}\ decays have a photon with an energy
above $30~\mbox{MeV}$, as shown in Fig.~\ref{introspettro},
corresponding to about 10\% of the center-of-mass energy.
These photons soften the momentum spectra of the electron
and pion that are actually detected in an experiment,
changing the geometrical acceptance and the distributions
of kinematic quantities often used to select or count
the number of signal events. If these effects are neglected, 
errors up to few percent can result in the measurement of a given 
branching ratio~\cite{ktevkl}~\cite{kloekl}.

\section{Bremsstrahlung and infrared divergences}

The main problem in simulating radiative decays
is the presence of infrared divergences:
the total decay width for single photon emission,
computed at any fixed order in $\alpha$, is infinite.
A finite value is obtained only by summing 
the decay widths for the real and virtual
processes calculated to the same order
in $\alpha$. 
As shown in Ref.~\cite{weinberg}, in the limit of soft photon
energy, we can ``re-sum'' the probabilities for multiple photon emission 
to all order in $\alpha$. The rate for the decay process $i\to f$
accompanied by any number of soft photons with total 
energy less than $E_{\gamma}$ is given by: 
\begin{equation}
  \label{weinbresult}
  \Gamma_{incl}(E_{\gamma})=
  \Gamma_{0} \left(\frac{E_{\gamma}}{\Lambda}\right)^{b}(1+O(b^{2})+O(E_{\gamma})). 
\end{equation}
Here $\Gamma_{0}$ is the unphysical decay width for the process 
$i\to f$ without final state photons, and $b$ is a function
of the particle momenta, is positive and of order $\alpha$; it is given by:
\begin{equation}
  \label{bondfactor}
  b = - \frac{1}{8 \pi^{2}} \sum_{m,n} \eta_{m} \eta_{n} e_{m} e_{n}
  \beta_{m n}^{-1} \ln \frac{1 + \beta_{m n}}{1 - \beta_{m n}},
\end{equation}
where $m$ and $n$ run over all the external particles,  $e_{n}$
is the charge of the particle $n$, $\eta=+1$ or $-1$ for an outgoing
or incoming particle, and $\beta_{m n}$ is the relative velocity
of the particles $n$ and $m$ in the rest frame of either:
\begin{equation}
  \label{beta}
  \beta_{m n} = \left[ 1 - \frac{m_{n}^{2} m_{m}^{2}}{(p_{n} \cdot p_{m})^{2}}
    \right]^{1/2} .
\end{equation}
$\Lambda$ is an energy cut-off that can be chosen as the mass $M$
of the decaying particle.

Differentiating $\Gamma_{incl}(E_{\gamma})$ with respect to $E_{\gamma}$,
we obtain an integrable differential distribution:
\begin{equation}
  \label{diffcrosfinit}
  \frac{d\Gamma_{incl}}{dE_{\gamma}} = 
  \Gamma_{0} b \frac{E_{\gamma}^{b-1}}{M^{b}}=
  \frac{d\Gamma_{Brem}}{dE_{\gamma}} 
  \left(\frac{E_{\gamma}}{M}\right)^{b},
\end{equation}
where we have neglected second order terms in $b$.
To $O(\alpha$) this can be identified with the single-photon
emission probability, and indeed, it can be written in terms of: 
\begin{equation}
  \label{cross}
  \frac{d\Gamma_{Brem}}{dE_{\gamma}}= 
  \Gamma_{0}\frac{b}{E_{\gamma}} . 
\end{equation}
The presence of the extra factor $(E_{\gamma}/M)^{b}$
ensures the integrability of Eq.~(\ref{diffcrosfinit})
in the limit $E_{\gamma}\to 0$.

In the derivation of Eq.~(\ref{weinbresult}), in Ref.~\cite{weinberg}, 
no explicit integration is required on the momenta of particles 
other than those of the photons. 
The result in Eq.~(\ref{diffcrosfinit}) can thus be
applied to differential decay withs: 
\begin{equation}
  \label{diffcrosfinit2}
  \frac{d\Gamma_{incl}}{dE_{\gamma}d\underline{\xi}} = 
  \frac{d\Gamma_{Brem}}{dE_{\gamma}d\underline{\xi}} 
  \left(\frac{E_{\gamma}}{M}\right)^{b(\underline{\xi})},
\end{equation}
where $\underline{\xi}$ represents the independent
kinematic variables of the decay process without photons,
and where:
\begin{equation}
  \label{cross2}
  \frac{d\Gamma_{Brem}}{dE_{\gamma}d\underline{\xi}}= 
  \frac{d\Gamma_{0}}{d\underline{\xi}}\frac{b(\underline{\xi})}{E_{\gamma}}. 
\end{equation}
Note that, while for two body decays $b$ is a constant,
for decays with more particles in the final state,
the velocities $\beta$ in Eq.~(\ref{beta}) and thus $b$, 
depend on the variables $\underline{\xi}$.

\section{MC simulation for radiative decays}

While for a complete MC simulation we need 
the decay width for all values of $E_{\gamma}$,
the relation in Eq.~(\ref{diffcrosfinit2}) is true for 
only soft-photon emission.
However, the value of the exponent $b$, which is of order $\alpha$, 
is about $0.01$. Hence, while $(E_{\gamma}/M)^{b}$
is a big correction for $E_{\gamma} \to 0$, its value is close to 1
when  $E_{\gamma} \to M$.
Therefore, if we use the complete differential decay width at order $\alpha$
for the emission of a photon, $d\Gamma_{Brem}/dE_{\gamma}d\underline{\xi}$,
instead of its approximation for low energies in Eq.~(\ref{cross2}),
the decay width in Eq.~(\ref{diffcrosfinit2}) 
represents a good approximation for the entire energy spectrum.

In summary, the ``recipe'' for writing the  amplitude
for the process $i \to f \gamma$ is the following:
\begin{itemize}
\item[1)] Calculate the $b$ factor using Eq.~(\ref{bondfactor}). 
\item[2)] Calculate the amplitude $M_{i\to f\gamma}$ at order $\alpha$.
\item[3)] Fix the divergence in the squared amplitude by multiplying  
  by $(E_{\gamma}/M)^{b}$: 
  \begin{equation}
    \left| M_{i\to f\gamma} \right|^{2}_{no IR} = 
    \left| M_{i\to f\gamma} \right|^{2} (E_{\gamma}/M)^{b} . 
  \end{equation}
\end{itemize}

\subsection{MC generators}

Following the recipe outlined in the previous section we have written 
the MC generators for the decays:
$K^{0}\to\pi\pi\gamma$, $K^{0}\to \pi e \nu\gamma$, 
$K^{0}\to \pi \mu \nu\gamma$, $K^{0}\to \pi^{+}\pi^{-}\pi^{0}\gamma$, 
$K^{\pm}\to\pi\pi\gamma$, $K^{\pm}\to e \nu\gamma$,
$K^{\pm}\to \mu \nu\gamma$,  $K^{\pm}\to \pi e \nu\gamma$, 
$K^{\pm}\to \pi \mu \nu\gamma$, and 
$K^{\pm}\to \pi^{\pm}\pi^{0}\pi^{0}\gamma$.
We obtained the amplitudes $M_{i\to f\gamma}$ at order $\alpha$  
mainly from Ref.~\cite{dafneh},
where they are calculated using chiral perturbation theory
at order $p^{2}$ and $p^{4}$. 
Concerning the semileptonic modes, for simplicity we have used only
the $p^{2}$ expression of the $M_{i\to f\gamma}$ amplitudes.
At this order, the form factors $f_{+}$ and $f_{-}$
are equal to $1$ and $0$ respectively.
In order to take into account the leading dependence on the
variable $t=(p_{K}-p{\pi})^2$ we have multiplied 
the amplitude by an overall factor $(1+\lambda_{+} t/M_{\pi+}^{2})$.
\footnote{We used the value $\lambda_{+}=0.03$.}  
The uncertainty related to this approximation is discussed
in the following. 

Large MC production in high energy experiments puts stringent 
limits on the time needed to generate one single event. This time
should not exceed the time needed to track the particles inside
the detector. In the KLOE experiment this time is about a few milliseconds.  
We used a combination of MC sampling techniques,
the {\em acceptance-rejection} (Von Neumann) method and the
{\em inverse transform} method~\cite{casual}, 
to reach this goal~\cite{notageneratori}. The average time for 
generating one event is a fraction of a millisecond.

We compared the fraction of events 
with a photon above an energy threshold predicted by the MC simulation,
with theoretical  expectations and experimental results, 
for several decay channels of the kaon.
For instance, the MC prediction for 
the fraction of $K^{0}\to\pi^{+}\pi^{-}\gamma$  in which a photon has energy 
greater than 20 MeV (50 MeV)
is equal to $7.00\times 10^{-3}$ ($2.54\times 10^{-3}$), in agreement
with the measured value $(7.10 \pm 0.22)\times 10^{-3}$ 
($(2.56\pm 0.09)\times 10^{-3}$),
and that from theoretical predictions $7.01\times 10^{-3}$ ($2.56\times 10^{-3}$)
for the $K_{S}$\footnote{These numbers
  refers only to the inner bremsstrahlung (IB) term.} 
(see Ref.~\cite{ramberg}).

Moreover, we calculated the ratio
\begin{equation}
\label{defineratio}
  R=\frac{\Gamma(K_{e3\gamma}, E_{\gamma}>E, \theta_{e\gamma}>\theta)}
  {\Gamma(K_{e3}(\gamma))} , 
\end{equation}
giving the fraction of decays $k\to \pi e \nu\gamma$ with a photon
with energy $E_{\gamma}$ above $E$ and angle between the electron 
and photon $\theta_{e\gamma}$ above $\theta$, 
for different values of the energy $E$ and angle $\theta$,
and we compared it with theoretical predictions and, when ever possible, with 
experimental values.
MC calculations of $R^{0}$ for \keg{0}\ decays are shown in 
Tab.~\ref{tabk0e3gmc}, while theoretical expectations
from Ref.~\cite{doncel} and 
Ref.~\cite{fearing}\footnote{We quote the values from Ref.~\cite{fearing},
obtained for $\lambda_{+}=0.03$, as in our simulation.} 
are shown in the top and middle part of Tab.~\ref{tabk0e3gtheo} 
respectively. Since the decay width in the denominator of Eq.~(\ref{defineratio})
is inclusive of the photon emission, following Ref.~\cite{ktevke3g}, 
we divided the predictions of Ref.~\cite{doncel} and~\cite{fearing}
by a factor $(1+\delta^{e}_{k})$, where $\delta^{e}_{k}$ is the
total electromagnetic correction extracted from  Ref.~\cite{cirigliano}.
Note that a different approach is used in Ref.~\cite{gasser}, where most
of the electromagnetic corrections are absorbed in $f_{+}(0)$ 
and therefore cancel out in the ratio $R^{0}$. 

The results for $R^{0}$
from a recent MC simulation, described in Ref.~\cite{tandre},
are shown in the lower part of Tab.~\ref{tabk0e3gtheo}.
Experimental results for $R^{0}$ have been recently published in 
Ref.~\cite{ktevke3g} by the KTeV Collaboration, 
for two values of the photon energy and angle:
\begin{eqnarray}
\nonumber
\frac{\Gamma(K_{e3\gamma}, E_{\gamma}>10~\mbox{MeV}, \theta_{e\gamma}>0^{\circ})}{\Gamma(K_{e3}(\gamma))} & = & (4.942\pm 0.062)\times 10^{-2},
\\
\nonumber
\frac{\Gamma(K_{e3\gamma}, E_{\gamma}>30~\mbox{MeV}, \theta_{e\gamma}>20^{\circ})}{\Gamma(K_{e3}(\gamma))} & = & (0.916\pm 0.017)\times 10^{-2},
\end{eqnarray}
and in Ref.~\cite{na48ke3g} by the NA48 Collaboration: 
\begin{eqnarray}
\nonumber
\frac{\Gamma(K_{e3\gamma}, E_{\gamma}>30~\mbox{MeV}, \theta_{e\gamma}>20^{\circ})}{\Gamma(K_{e3}(\gamma))} & = & (0.964\pm 0.008^{+0.011}_{-0.009})\times 10^{-2}.
\end{eqnarray}

MC calculations and theoretical expectations
of $R^{\pm}$ for \keg{\pm}\ decays are shown in 
Tab.~\ref{tabkce3gmc} and in Tab.~\ref{tabkce3gtheo}.
Since the radiative correction  $\delta^{e}_{k}$ for $\Gamma(K^{\pm}_{e3})$ 
is compatible with zero, $\delta^{e}_{k}=+0.06(20)\%$,
it has been neglected.

In Tab.~\ref{tabk0e3gmc} and Tab.~\ref{tabkce3gmc}
the first error is statistical while the second is the systematic one.
We computed the systematic error by comparing the difference
between the branching ratios calculated at order $O(p^2)$ and 
at order $O(p^4)$ in Ref.~\cite{dafneh},
with the variations of $R$ we observed in the simulation when 
the term $(1+\lambda_{+} t/M_{\pi+}^{2})$ is included, or not, 
in the decay amplitude.
In Ref.~\cite{dafneh} the branching ratios evaluated 
for $E_{\gamma}>30~\mbox{MeV}$ and $\theta_{\gamma}>20^{\circ}$
increase by about 6\% in the $O(p^4)$ calculation, while 
the variations of $R$ in the simulation are about 3\%. Hence, 
we used the variations observed in the simulation for each of 
$E_{\gamma}$ and $\theta_{\gamma}$ as systematic errors. 
Moreover, we changed the value of the cut-off $\Lambda$ 
from $M_{K}$ to $M_{K}/2$ to check the stability of the results,
resulting in negligible variations of the order of the statistical
errors shown in the two tables.

For both $R^{0}$ and $R^{\pm}$ the absolute differences between the MC
simulation and theoretical predictions are below $5 \times 10^{-4}$,
and below the quoted systematic errors.
Moreover, such errors are smaller than the relative errors on 
measured branching ratios of kaon decays~\cite{ktevkl}~\cite{kloekl}~\cite{na48kl}.

\begin{table}
  \begin{center}
    \begin{small}
    \begin{tabular}{|c|c|c|c|c|}\hline
      $\theta$/E(MeV) & 10 & 20  & 30  & 40 \\ \hline
      $0^{o}$     & $4.908(7)(56)$ & $3.252(6)(47)$ & 
      $2.364(5)(43)$ & $1.784(4)(38)$ \\ \hline
      $10^{o}$    & $2.450(5)(38)$ & $1.657(4)(36)$ & $1.223(3)(31)$ & $0.937(3)(28)$ \\ \hline
      $20^{o}$    & $1.864(4)(33)$ & $1.262(4)(30)$ & $0.933(3)(28)$ & $0.716(3)(24)$ \\ \hline 
      $30^{o}$    & $1.516(4)(30)$ & $1.028(3)(28)$ & $0.760(3)(25)$ & $0.583(2)(22)$ \\ \hline 
      $40^{o}$    & $1.264(4)(27)$ & $0.859(3)(25)$ & $0.636(3)(23)$ & $0.488(2)(21)$ \\ \hline 
      $50^{o}$    & $1.066(3)(24)$ & $0.726(3)(23)$ & $0.538(2)(20)$ & $0.414(2)(19)$ \\ \hline  
    \end{tabular} 
    \end{small}
  \end{center}
  \caption{Ratios $R^{0}\times10^{2}$ for $K^{0}e3$ decays obtained in the MC simulation
    with $10^{7}$ events. The first error is statistical while the second is systematic.}
  \label{tabk0e3gmc}
\end{table}

\begin{table}
  \begin{center}
    \begin{tabular}{|c|c|c|c|c|}
      \hline
      Ref.~\cite{doncel} $\theta$/E(MeV) & 10 & 20  & 30  & 40  \\ \hline
      $0^{o}$     & $4.94$ & $3.25$ & $2.35$ & $1.77$ \\ \hline
      $10^{o}$    & $2.48$ & $1.67$ & $1.24$ & $0.95$ \\ \hline
      $20^{o}$    & $1.90$ & $1.29$ & $0.95$ & $0.73$ \\ \hline 
      $30^{o}$    & $1.55$ & $1.05$ & $0.78$ & $0.60$ \\ \hline 
      $40^{o}$    & $1.29$ & $0.85$ & $0.66$ & $0.50$ \\ \hline 
      $50^{o}$    & $1.10$ & $0.75$ & $0.56$ & $0.43$ \\ 
      \hline\hline  
      Ref.~\cite{fearing} $\theta$/E(MeV) & 10 & 20  & 30  & 40 \\ \hline
      $0^{o}$    & $-$    & $3.27$ & $2.37$ & $1.78$  \\ 
      \hline\hline
      Ref.~\cite{tandre}  $\theta$/E(MeV) & 10 & 20  & 30  & 40 \\ \hline
      $0^{o}$     & $4.93(6)$ & $-$  & $2.36(3)$& $-$  \\ \hline
      $20^{o}$    & $1.89(2)$ & $-$ & $0.96(1)$ & $-$ \\ \hline 
    \end{tabular} 
  \end{center}
  \caption{Ratios $R^{0}\times10^{2}$ for $K^{0}e3$ decays: 
    (Top) Values listed in Ref.~\cite{doncel};
    (Middle) Values obtained from Ref.~\cite{fearing}; (Bottom) 
    Values from the MC simulation in Ref.~\cite{tandre}.
    The values from Ref.~\cite{doncel} and Ref.~\cite{fearing} 
    have been multiplied by $(1+\delta^{e}_{k})^{-1}$ where $\delta^{e}_{k}=1.04(20)\%$ 
    from Ref.~\cite{cirigliano}.}
  \label{tabk0e3gtheo}
\end{table}

\begin{table}
  \begin{center}
    \begin{small}
      \begin{tabular}{|c|c|c|c|c|}\hline
	$\theta$/E(MeV) & 10 & 20  & 30  & 40 \\ \hline
	$0^{o}$     & $4.223(6)(34)$ & $2.825(5)(28)$     & $2.069(5)(27)$ & $1.572(4)(25)$ \\ \hline
	$10^{o}$    & $1.781(4)(18)$ & $1.238(3)(17)$ & $0.936(3)(17)$ & $0.732(3)(15)$ \\ \hline
	$20^{o}$    & $1.211(3)(16)$ & $0.854(3)(15)$ & $0.652(3)(14)$ & $0.515(2)(13)$ \\ \hline 
	$30^{o}$    & $0.881(3)(15)$ & $0.630(3)(14)$ & $0.488(2)(13)$ & $0.388(2)(12)$ \\ \hline 
	$40^{o}$    & $0.659(3)(14)$ & $0.477(2)(13)$ & $0.373(2)(11)$ & $0.300(2)(10)$ \\ \hline 
	$50^{o}$    & $0.493(2)(12)$ & $0.363(2)(11)$ & $0.289(2)(11)$ & $0.234(2)(10)$ \\ \hline  
      \end{tabular} 
    \end{small}
  \end{center}
  \caption{Ratios $R^{\pm}\times10^{2}$ for $K^{\pm}e3$ decays as obtained in the MC simulation with $10^{7}$ events. The first error is statistical while the second is systematic.}
  \label{tabkce3gmc}
\end{table}

\begin{table}
  \begin{center}
    \begin{tabular}{|c|c|c|c|c|}\hline
      Ref.~\cite{doncel} $\theta$/E(MeV) & 10 & 20  & 30  & 40 \\ \hline
      $0^{o}$     & $4.24$ & $2.80$ & $2.03$ & $1.53$ \\ \hline
      $10^{o}$    & $1.78$ & $1.22$ & $0.91$ & $0.71$ \\ \hline
      $20^{o}$    & $1.20$ & $0.83$ & $0.63$ & $0.50$ \\ \hline 
      $30^{o}$    & $0.87$ & $0.61$ & $0.47$ & $0.37$ \\ \hline 
      $40^{o}$    & $0.65$ & $0.46$ & $0.36$ & $0.29$ \\ \hline 
      $50^{o}$    & $0.48$ & $0.35$ & $0.27$ & $0.22$ \\ \hline\hline  
      Ref.~\cite{fearing} $\theta$/E(MeV) & 10 &20  & 30  & 40 \\ \hline
      $0^{o}$     & $-$ & $2.82$ & $2.04$ & $1.54$ \\ \hline  
    \end{tabular} 
  \end{center}
  \caption{Ratios $R^{\pm}\times10^{2}$ for $K^{\pm}e3$ decays: 
    (Top) Values listed in Ref.~\cite{doncel};
    (Bottom) Values obtained from Ref.~\cite{fearing}.
    The radiative correction $\delta^{e}_{k}=+0.06(20)\%$
    is compatible with zero and has been neglected.}
  \label{tabkce3gtheo}
\end{table}

\section{Conclusions}

We overcome the problem of infinite probabilities in radiative processes 
by extending the soft-photon approximation of Ref~\cite{weinberg}
to the whole energy range. The spectra produced with MC generators developed
with this technique, agree well with other theoretical calculations
and with available experimental data. The systematic error could be further 
reduced by using full $O(p^4)$ calculation for the amplitudes, or even adding
the $O(p^6)$ results recently published in Ref.~\cite{gasser}.

We used MC sampling techniques to reduce the time needed to 
generate one event below 1 ms. The MC generators, routines written in Fortran, 
have been included in the official KLOE library and have
been used for large MC productions. Such routines are available on request
to the author (claudio.gatti@lnf.infn.it).    

\section{Acknowledgments}
I would like to thank Gino Isidori and Mario Antonelli 
for many useful hints and discussions.

\end{document}